\documentstyle[12pt]{article}

\documentstyle{titlepage}
\title{ON THE POSITION OPERATOR FOR MASSLESS PARTICLES}
\author{\bf ALI\ SHOJAI$^*$\ \&\ MEHDI\ GOLSHANI$^{**}$\\
Department of Physics, Sharif University of Technology\\P.O.Box 11365-9161 Tehran, IRAN\\
and\\Institute for Studies in Theoretical Physics and Mathematics,\\P.O.Box 19395-5531, Tehran, IRAN\\
$^*$Email: SHOJAI@PHYSICS.IPM.AC.IR\\
$^{**}$Fax: 98-21-8036317\\}
\date{}
\begin{document}
\begin{bf}
\maketitle
\vspace{1cm}
\begin{center}
{\Large ON THE POSITION OPERATOR FOR MASSLESS PARTICLES}\\
{\bf A. Shojai \& M. Golshani}
\end{center}
\vspace{0.5cm}
\begin{center}
{\bf ABSTRACT}
\end{center}
{\it It is always stated that the position operator for massless particles has
non-comutting components. It is
shown that the reason is that the commutation relations between
coordinates and momenta differs for massive and massless particles.
The correct one for massless particles and a
position operator with commuting components are derived.}
\vspace{1.5cm}
\begin{center}
{\large \bf \S1. INTRODUCTION AND SURVEY}
\end{center}
\par
The notion of position operator has its roots in the early days of the birth of quantum mechanics. Although in the Copenhagen interpretation of quantum mechanics, the concept of position, and therefore
path of the particle, is meaningless, nevertheless there must exist an operator called position operator having the property that its expectation value in the classical limit would behave classically. In other words, any macroscopic
object has position. In quantum mechanics, one deals with {\it elementary systems\/} which means any system whose state has a definite transformation under Poincare group (or under Gallileo group in the non-relativistic case). An {\it elementary
particle\/}, then, can be defined as an elementary system which has no constituents. In this way, electron is an elementary particle while Hydrogen atom is an elementary system only. In dealing with elementary
systems one works only with generators of Poincare group as physical observables rather than the {\it position\/} of the system. Clearly it is natural to search for a {\it position operator\/} as an observable whose eigenvalues are the possible positio
of an elementary particle or the center of mass position of an elementary system. Unfortunately when one restricts himself to the positive energy manifold, the operator $i\vec{\nabla}_p$ is no longer hermitian.
\par
The problem of finding the position operator in the framework of nonrelativistic quantum mechanics, where the symmetry of space-time is
the Gallileo group, is simple$.^{[1,3]}$ Serious work on relativistic case began after the works of
Pryce and Newton-Wigner$.^{[2]}$ They found a position operator, which we call it Pryce-Newton-Wigner operator, having the foregoing property. Until now, a lot of theoretical works has been done on this operator$.^{[3]}$
\par
In spite of these investigations
concerning the position operator, the following problem observed by Pryce and Newton-Wigner, is still unsolved. When one tries to write down the position operator for massless particles with non-zero helicity (e.g.
photons), one encounters inconsistency.
Technically speaking, one is not able to write a position operator having commuting components for such particles. This is a serious problem, as photon would not be localizable. If you measure some component of the
photon's position, its other components could not be determined precisely, as Heisenberg's uncertainty principle dictates. It can be shown that the localizability problem is related to causality$.^{[4]}$
\par
Some people have tried to overcome the problem by rejecting or weakening the Newton-Wigner postulates for
the derivation of the position operator$.^{[5]}$ They assume that the probability of finding the particle in a volume $V$ consisting of volumes $V_1$ and
$V_2$ with empty intersection is not equal to the sum of the probabilities of finding the particle in $V_1$ and $V_2$. This is not a physically reasonable assumption, and still has the causality problem.
\par
In spite of the lack of localizability for massless particles, it has been shown$^{[6]}$ that by
defining precisely the concept of localizability for massless particles,
there exist localized wavefunctions with any desired accuracy, in accordance with the
experimental facts.
\par
In this work, we shall show that if one proposes that for massless particles
the canonical commutation relation is not the standard one, it can be shown that
one arrives at a position operator for massless particles which has commuting components.
In the appendix we present a classical argument in favour of the new commutation relation.
\par
Before doing so it is instructive to review breifly the procedure of constructing position operator for relativistic massive particles.
In quantum mechanics, observables are identified by hermitian linear operators with their eigenvalues as allowed results of any measurement of that observable. The essential problem is: what are the observables and their corresponding operators.
This can be answered in a formal way. Events occure in space and time and thus it is natural
to look for the symmetries of the space-time, which is according to the special theory of relativity, the Poincare group.
\par
The Poincare group consists of space and time translations, rotations and boosts generated by hermitian operators $\vec P$, $H$, $\vec J$ and $\vec K$ respectively. To these operations space inversion with unitary operator $\Pi$ and time reversal
with antiunitary operator $\cal T$ must be added. All of our knowledge about these operators are their commutation relations:\\
\begin{equation}
\left \{ \begin{array}{llll} [P_i,P_j]=0 & [P_i,H]=0 & [J_i,J_j]=i\epsilon_{ijk}J_k & [K_i,K_j]=-i\epsilon_{ijk}J_k\cr
                             [J_i,P_j]=i\epsilon_{ijk}P_k & [J_i,K_j]=i\epsilon_{ijk}K_k & [J_i,H]=0 & [K_i,P_j]=i\delta_{ij}H\cr
                             [K_i,H]=iP_i & {\Pi}^2 =1 & {\cal T}^2 =1 & [{\Pi},{\cal T}]=0\cr
                             {\Pi} \vec P {\Pi}=- \vec P & {\Pi} H {\Pi}=H & {\Pi} \vec J {\Pi}= \vec J & {\Pi} \vec K {\Pi}=- \vec K\cr
                             {\cal T} \vec P {\cal T}=- \vec P & {\cal T} H {\cal T}=H & {\cal T} \vec J {\cal T}=- \vec J & {\cal T} \vec K {\cal T}= \vec K \end{array} \right .
\end{equation}
with clear physical meanings. Note that ${\cal T}$ is antiunitary, i.e. acting on any function leads to its complex conjugate:
\begin{equation}
{\cal T}f=f^*
\end{equation}
Irreducible representations of the Poincare group which are identified as particles according to Wigner, can be constructed using the Casimir operators:
\begin{equation}
{\cal C}_1=H^2-P^2
\end{equation}
\begin{equation}
{\cal C}_2=(\vec P \cdot \vec J)^2 - (H \vec J + \vec P \times \vec K)^2
\end{equation}
Now following Foldy let$^{[7]}$
\begin{equation}
\vec J=\vec Q \times \vec P + \vec S
\end{equation}
\begin{equation}
\vec K=\frac{1}{2} (H \vec Q+\vec Q H)+H^{-1}\vec P \times \vec S -t \vec P
\end{equation}
where $\vec Q$ must be identified as the position operator of the particle, $\vec L=\vec Q \times \vec P$ as the orbital angular momentum and $\vec S$ as the spin. Using the canonical commutation relation:
\begin{equation}
[Q_i,P_j]=i\delta_{ij}
\end{equation}
and after some algebra, one can show:
\begin{equation}
{\cal C}_2=-m^2S^2
\end{equation}
\begin{equation}
[S_i,S_j]=i\epsilon_{ijk}S_k
\end{equation}
the last relation enables one to interpret $\vec S$ as spin. From these relations the position operator can be derived:
\begin{equation}
\vec Q=H^{-1} (\vec K + t\vec P -\frac{i}{2}H^{-1}\vec P)-m^{-1}H^{-1}(H+m)^{-1}\vec P \times (H \vec J+\vec P \times \vec K)
\end{equation}
This is the Pryce-Newton-Wigner position operator. Note that this is meaningless in the limit $m\rightarrow 0$. Its time derivative is the velocity:
\begin{equation}
\frac{d\vec Q}{dt}=i[H,\vec Q]=H^{-1}\vec P
\end{equation}
It is a vector:
\begin{equation}
[J_i,Q_j]=i\epsilon_{ijk}Q_k
\end{equation}
\begin{equation}
{\Pi}\vec Q {\Pi} =-\vec Q
\end{equation}
and under boosts:
\begin{equation}
[K_i,Q_j]=-iH^{-1}P_iQ_j
\end{equation}
and time reversal:
\begin{equation}
{\cal T}\vec Q {\cal T} =\vec Q
\end{equation}
It can be shown that this position operator is unique up to canonical transformations$.^{[1]}$\\
\\
\begin{center}
{\large \bf \S2. POSITION OPERATOR FOR MASSLESS PARTICLES}
\end{center}
\par
For massless particles the Pryce-Newton-Wigner position operator does not work
as it can be seen from the fact that
in the $m\rightarrow 0$ limit, it does not have a good behaviour.
In fact if one starts with massless representations of Poincare group the
position operator obtained has not commuting components if one uses the
canonical commutation relation (7). Thus this leads to non-localizability
(which is equal to the lack of causality).
It can be seen that this is because the correct commutation relation for
position and momentum is not used.
It can be argued {\it heuristically\/} that the correct one is as follows
\begin{equation}
[Q_i,P_j]=iH^{-2}P_iP_j
\end{equation}
instead of (7).
In this section we shall show that using this relation one will arrive at a
position operator with commuting components.
\par
For massless representations one has:
\begin{equation}
H^2=P^2
\end{equation}
\begin{equation}
\vec P \cdot \vec J= H \Sigma
\end{equation}
\begin{equation}
H\vec J + \vec P \times \vec K = \vec P \Sigma
\end{equation}
where $\Sigma$ is the helicity operator having eigenvalues $\pm h$ (for photon $h=1$). It can be shown that these equations leads to a non-commuting position operator except in the case of zero helicity$.^{[3]}$
\par
As it is shown in the previous section for massless particles the commutation relation (7) must be replaced by one given in equation (16).
Thus the problem is finding an operator satisfying relations (11)-(16). This operator
can be only of the forms $\vec f(\vec P) \cdot \vec K$, $g(H)\vec P \times \vec J$ and $h(H)\vec P \times (\vec P \times \vec K)$ because of the character of position
operator under space inversion and time reversal. The above equations may be used to solve for $f$, $g$ and $h$. The final result after symmetrization is as follows:
\begin{equation}
\vec Q= \frac{1}{2}(H^{-3}\vec P (\vec P \cdot \vec K)+ (\vec K \cdot \vec P) \vec P H^{-3}) +t H^{-1} \vec P
\end{equation}
This position operator has commuting components and all other commutation
relations are correct. Now, we should be careful about
two points: First; since we find the complete solution of commutation relations,
our position operator is unique up to a
canonical transformation (which leaves equation (16) unchanged, not equation
(7)). Second; according to equation (16) our position operator is not an
ordinary {\it vector\/} under translations. It is not a free vector, i.e.
when one
translate the reference frame the position operator of massless particles
does not move rigidly. (This is apparent from the fact that $\vec L =0$)
\par
This new position operator has at least two new intresting results which we shall
discuss below. First, since the standard position--momentum commutation relation
is changed for the massless particles, the uncertainty relation for position and
momentum may differ from the well known one. It is a standard result of quantum mechanics
that
\begin{equation}
(\Delta Q_i)(\Delta P_j) \ge \frac{1}{2}|<[Q_i, P_j]>| = \frac{\hbar}{2} |<H^{-2}P_i P_j>|
\end{equation}
where we have recovered the $\hbar$ factor.
\par
In order to calculate the right hand side of this relation we introduce
the momentum eigenstates as
\begin{equation}
P_i|\vec{k}>=k_i|\vec{k}>
\end{equation}
and write the general state of the system as follows
\begin{equation}
|\alpha>=\int d^3k S(\vec{k})|\vec{k}>
\end{equation}
It can be easily seen that
\begin{equation}
<H^{-2}P_iP_j>=\int d^3k |S(\vec{k})|^2 \frac{k_i k_j}{k^2}
\end{equation}
In the case in which $S$ is only a function of the lenght of $\vec{k}$,
using the orthonormality condition of the state vector we have
\begin{equation}
<H^{-2}P_iP_j>=\frac{1}{3}\delta _{ij}
\end{equation}
so
\begin{equation}
(\Delta Q_i)(\Delta P_j) \ge \frac{1}{6}\hbar \delta _{ij}
\end{equation}
In the general case where $S$ depends on the direction of $\vec{k}$ (i.e. when
there is a preffered direction $\vec{k}_0$) like $S(\vec{k})\sim exp(-\alpha (\vec{k}-\vec{k}_0)^2)$
the uncertainty relation for position and momentum reads as
\begin{equation}
(\Delta Q_i)(\Delta P_j) \ge \hbar (A \delta _{ij} +B k_{0i} k_{0j})
\end{equation}
where
\[ A=\frac{1}{6}+\ terms\ involving\ k_0\]
\[ B=0+\ terms\ involving\ k_0\]
Thus our new position operator suggests a new uncertainty relation for position and
momentum. The experimental consequences of this new relation can be in principle,
verified for  massless particles with a wavefunction which peaks at a very high
momentum, for example. In such a case the role of the second term at the right hand side of
the relation (27) is important.
\par
The second important result of our new position operator for massless particles
is about its eigenfunctions. To simplify the calculations, we work in the momentum
representation where
\begin{equation}
\vec{K}=\frac{1}{2}i(\frac{\partial}{\partial \vec{P}}H+H\frac{\partial}{\partial \vec{P}})-H^{-1}\vec{S} \times \vec{P}
\end{equation}
and consider a zero helicity massless particle. The Shcrodinger picture eigenvalue
problem for the position operator is then
\begin{equation}
iP^{-2}\vec{P}\vec{P}\cdot \frac{\partial \Phi_{\vec{q}}(\vec{P})}{\partial \vec{P}}+iP^{-2}\vec{P}\Phi_{\vec{q}}(\vec{P})=\vec{q} \Phi_{\vec{q}}(\vec{P})
\end{equation}
where $\Phi_{\vec{q}}(\vec{P})$ is the position eigenfunction in the momentum representation with the eigenvalue $\vec{q}$.
The form of this equation suggests that the wavefunction is nonzero only when
$\vec{P}$ and $\vec{q}$ are parrallel. So we set
\begin{equation}
\Phi_{\vec{q}}(\vec{P})=\Phi^{(0)}_{\vec{q}}(P) \delta \left ( \frac{\vec{P}\cdot \vec{q}}{Pq}-1 \right )
\end{equation}
Inserting this relation in (29) one arrives at
\begin{equation}
\frac{d \Phi^{(0)}_{\vec{q}}(P)}{dP}=\left ( -iq -\frac{1}{P}\right ) \Phi^{(0)}_{\vec{q}}(P)
\end{equation}
which can be easily solved. The wavefunction is thus
\begin{equation}
\Phi_{\vec{q}}(\vec{P})=\frac{N}{P}e^{-iPq} \delta \left ( \frac{\vec{P}\cdot \vec{q}}{Pq}-1 \right )
\end{equation}
The $\vec{x}$-representation position eigenfunctions can be obtained via Fourier transformation
\begin{equation}
\Psi_{\vec{q}}(\vec{x})=\int \frac{d^3P}{P}\Phi_{\vec{q}}(\vec{P}) e^{-i\vec{P}\cdot \vec{x}}=\frac{N}{2}\delta \left ( \frac{\vec{x}\cdot \vec{q}}{q}-q \right )
\end{equation}
That is our position operator for massless particles has the peculiar property
that is delta function in the direction of its eigenvalue and is constant
in the direction prependicular to the eigenvalue.
\\

\begin{center}
{\large \bf \S3. CONCLUSION}
\end{center}
\par
It is shown that on the basis of classical arguments one is forced to propose a new commutation relation between
position and momentum for massless particles.
Using the new commutation relation (16) one arrives at a position operator for massless particles which has commuting components.
The effect of the new commutation relation on the position-momentum uncertainty relation is investigated.
Also the localized states, i.e. the eigenfunctions of this new position operator
are derived.
\newline
\begin{center}
{\large \bf APPENDIX}
\end{center}
\par
In this appendix we presenta classical reasoning in favour of the relation (16).
Consider a classical system with coordinate $\vec{Q}$, energy $H$ and momentum $\vec{P}$.
According to the well known results of classical mechanics, translation of
the reference frame by $\vec{\epsilon}$ affects the coordinates as
\[ Q'_i= Q_i+\epsilon_j\{P_j,Q_i\} \]
where $\{,\}$ represents  Poisson brackets. If the standard Poisson bracket between coordinates and momenta
is satisfied
\[\{Q_i,P_j\}=\delta _{ij} \]
we have
\[\vec{Q}'=\vec{Q}+\vec{\epsilon} \]
which reads as: {\it translation of reference frame is equal to translation of the particle\/}. Why these two operations are equal?
This is because for a massive particle one can always transform to the rest frame of the particle, in which the particle
is attached to the space-time.
\par
Now the difficulty for massless particles is apparent. Any massless particle
must move with unit velocity (for such particles $H^2=P^2$ and thus
$u^2=H^{-2}P^2 =1$) and it is a well-known result of Poincare transformations
that there is no rest frame for
such particles -- they move with unit velocity in any reference frame.
Thus one cannot use the standard Poisson brackets  for coordinates and momenta.
Let us see what is the correct one for  massless particles.
\par
Let the velocity of the particle be $\vec u$ with $u^2=1$ and suppose we want
to calculate ${P_1,Q_2}$. So we choose $\vec \epsilon=\epsilon
\hat {e_1}$ and transform to a frame in which $u_1'=0$. For simplicity we assume that the translation of reference frame
is dynamic, i.e.:
\[ \epsilon=u_1t'\]
During time $t'$,
the particle moves in $\hat {e_2}$ direction by $u_2't'$ which when
transformed to the initial frame is equal to $u_2t\gamma^{-2}(u_1)$.
From this amount $u_2t$ must be subtracted because we assume that the
translation to be dynamic. The
net change in $Q_2$ is:
\[ u_2t(1-u_1^2)-u_2t=-u_1^2u_2t=-u_1u_2\epsilon \]
so  we conclude that $\epsilon\{P_1,Q_2\}=-\epsilon u_1 u_2$ or in general:
\[ \{P_j,Q_i\}=H^{-2}P_iP_j \]
The quantum mechanical analogous of this relation can be achieved via Dirac's
canonical quantization rule $\{,\}\rightarrow -i[,]$ as
\[ [Q_i,P_j]=iH^{-2}P_iP_j \]
This equation is the analogous to equation (7) and must be used for massless particles.
It is worthwhile to note that it is covariant (i.e. is compatible with
equations (1)) and thus it is independent of the way it is constructed.
\\
\begin{center}
{\large \bf REFERENCES}
\end{center}
[1]-T.F. Jordan, J. Math. Phys., {\bf 21}(8), 2028, (1980).
\newline
[2]-M.H.L. Pryce, Proc. Roy. Soc. {\bf 195A}, 62 (1948); T.D. Newton and E.P. Wigner, Rev. Mod. Phys. {\bf 21}, 400 (1949).
\newline
[3]-See H. Bacry, {\it Localizability and Space in Quantum Physics\/}, Springer-Verlag (1988); and Z.K. Silagadze, Preprint SLAC-PUB-5754 Rev., (August 1993) for a complete bibilography.
\newline
[4]-J. Mourad and R. Omnes, Preprint LPTHE 92/41, (July 1992).
\newline
[5]-M. Comi, Il Nuovo Cimento, {\bf 56A}, No.3, 299, (April 1980); and references therein.
\newline
[6]-J. Mourad, Phys. Lett. A, {\bf 182}, 319, (1993).
\newline
[7]-L.L. Foldy, Phys. Rev., {\bf 102}, No.2, (1956).
\end{bf}
\end{document}